\documentclass[aps,prd,twocolumn,psfig,groupedaddress]{revtex4}

\begin{document}

\input psfig.sty


\title{Numerically generated quasi-equilibrium orbits of black holes: 
Circular or eccentric?}


\author{Thierry Mora }
\email[]{tmora@clipper.ens.fr}
\affiliation{\`Ecole Normale Sup\'erieure, D\'epartement de Physique,
\\23 Rue Lhomond, 75231 Paris Cedex 05, France}
\affiliation{McDonnell Center for the Space Sciences, Department of
Physics, \\Washington University, St. Louis, Missouri 63130}
\author{Clifford M. Will}
\email[]{cmw@wuphys.wustl.edu}
\homepage[]{wugrav.wustl.edu/People/CLIFF}
\affiliation{McDonnell Center for the Space Sciences, Department of
Physics, \\Washington University, St. Louis, Missouri 63130}


\date{\today}

\begin{abstract}
We make a comparison between results from 
numerically generated, quasi-equilibrium
configurations of compact binary systems of black
holes in close orbits, and results from the post-Newtonian approximation.  
The post-Newtonian results are accurate through third PN order
($O(v/c)^6$ beyond Newtonian gravity), and include 
rotational and spin-orbit effects, but are generalized to permit orbits of
non-zero eccentricity.  
Both treatments ignore gravitational radiation reaction.
The energy $E$ and angular momentum $J$ 
of a given configuration are compared between the two
methods as a function
of the orbital angular frequency $\Omega$.  
For small $\Omega$, corresponding to orbital separations a factor of
two larger than that of the innermost stable orbit, 
we find that, if the orbit is permitted to be slightly eccentric, with
$e$ ranging from $\approx 0.03$ to $\approx 0.05$,
and with the two objects
initially located at the orbital apocenter (maximum separation), our
PN formulae give much better fits to the numerically generated data
than do any circular-orbit PN methods, including various ``effective
one-body'' resummation
techniques.
We speculate that the
approximations made in solving the initial value equations of general
relativity numerically may introduce a spurious eccentricity into the
orbits.  
\end{abstract}

\pacs{}

\maketitle

\section{Introduction and Summary \label{intro}}
%
%
%
The late stage of inspiral of binary systems of neutron stars or black
holes is of great current interest, both as a challenge for numerical
relativity, and as a possible source of gravitational waves detectable
by laser interferometric antennas.  Because this stage, corresponding
to the final few orbits and ultimate merger of the two objects into
one, is highly dynamical and involves strong gravitational fields, it
must be handled by numerical relativity, which attempts to solve the
full Einstein equations on computers (see Refs. \cite{cook,seidel} for
reviews).

The early stage of inspiral can be handled accurately using
post-Newtonian techniques, which involve an expansion of solutions of
Einstein's equations in powers of $\epsilon \sim (v/c)^2 \sim
Gm/rc^2$, where $v$, $m$, and $r$ are typical velocity, mass and
separation in the system, respectively.  By expanding to very high
powers of $\epsilon$, increasingly accurate formulae can be derived to
describe both the orbital motion and the gravitational waveform.
Currently, results accurate through 3.5PN order ($O(\epsilon^{7/2})$
beyond Newtonian gravity) are known 
\cite{jaraschafer98,jaraschafer99,djs00,bf00,bf01,abf01,dire2}.

An important issue in understanding the full inspiral of compact
binaries is how to connect the PN regime to the numerical regime.
This is a non-trivial issue because the PN approximation gets worse
the smaller the separation between the bodies.  
On the other hand,
because of limited computational resources, numerical simulations
cannot always be started with separations sufficiently large to
overlap the PN regime where it is believed to be reliable.  

Numerical simulations of compact binary inspiral start with a solution
of the initial value (I-V) equations of Einstein's theory; these
provide the initial data for the evolution equations (some simulations
\cite{ggb02} solve in addition
one of the six dynamical field equations).  The initial
state is assumed to consist of two compact objects (neutron stars or
black holes) in an initially circular orbit.  
The circular-orbit condition is
imposed by demanding that $dr/dt = 0$ initially, where $r$ is a
measure of the orbital separation.  More precisely, the system is
presumed to have an initial ``helical Killing vector'' (HKV), which
corresponds to a kind of rigid rotation of the binary system.  This
amounts to ignoring initially
the effects of gravitational radiation reaction.  
It also implies that the black holes
are co-rotating, a condition which is astrophysically unlikely,
albeit computationally advantageous.
To make the problem tractable
numerically, it is also generally assumed that the spatial metric is
conformally flat.  This approximation is usually justified by the
neglect of radiation reaction in the initial state.  
For black hole binaries, suitable horizon boundary conditions must be
imposed, while for neutron star binaries, hydrodynamical equations and
an equation of state must be provided.  

One important product of these initial value solutions is a
relationship between the energy $E$ and angular momentum $J$
of the system as measured at infinity, and the orbital
frequency $\Omega$.  
As all  quantities 
are well-defined and gauge invariant, they are useful
variables for making comparisons with PN methods.  

We have developed a formula for $E(\Omega)$ and $J(\Omega)$
using PN methods.  Our analytic formula includes point-mass terms
through 3PN order, but ignores radiation reaction.  
It also includes rotational energy and
spin-orbit terms for the case in which the bodies are rotating.  
For black holes, tidal
effects can be ignored.  
In contrast to previous work \cite{luc02,dgg02}, our
formulae apply to general eccentric orbits, not just to circular orbits.

We then compare this formula with
HKV numerical solutions for corotating binary black holes obtained by
Grandcl\'ement et al. \cite{ggb02}, 
for the regime where the black holes are 
separated from the location of the innermost circular orbit by a
factor of around two, where PN
results might be expected to work well ($Gm/rc^2 \sim 0.1$). 
We find the following two results:

\noindent
1.  When we assume circular PN orbits, our PN formulae for 
$E(\Omega)$ and $J(\Omega)$ agree well with other PN methods, including those
using resummation or Pad\'e techniques.  However all PN methods 
consistently and systematically underestimate the binding energy and
overestimate the angular momentum, compared
to
the values derived from the numerical HKV 
simulations, by amounts that are larger
than the spread among the PN methods.

\noindent
2.
When we relax the assumption of a circular orbit and demand only
that $dr/dt =0$, our PN formula agrees extremely well with the numerical 
data.  In this case the system is found to be initially at the
apocenter of a slightly eccentric orbit.  For values of $Gm \Omega/c^3 $
ranging from 0.03 to 0.06, corresponding to orbital $v/c$ between 0.3 and 0.4, 
or orbital separation between 10 and 6 $Gm/c^2$, 
perfect agreement with the binding energy can be obtained
with eccentricities that
range from 0.03 to 0.05.

If this apparent eccentricity is a real effect, 
there are two possible explanations.
One is that
the imposition of a helical Killing vector, which is equivalent to
assuming a stationary binary star configuration as seen in a rotating
frame, is not sufficient to guarantee a circular orbit in the
relativistic regime.  In Newtonian
gravity, this assumption would be 
equivalent to imposing $dr/dt = 0$, which only puts
the orbit at a turning point.  An additional condition $d^2r/dt^2 =0$
is needed to enforce a circular orbit.  But this is a dynamical
statement, which is outside the realm of the I-V equations of
Einstein's theory.  It may be that, in solving one of the dynamical
field equations (the trace of the equation for $\partial
K^{ij}/\partial t$) in addition to the four I-V equations,
Grandcl\'ement et al. take care
of this automatically. 

The more likely possibility is that the approximations made in most numerical
solutions, the main one being that the spatial metric is conformally
flat, somehow introduce a systematic eccentricity into the orbit.

At present, these results can only be considered a hint of a possible 
eccentricity,
however.
It is entirely possible that circular PN orbits {\it are} consistent with
the HKV results within the errors of the numerical simulations.  This can
only be decided if and when
the numerical groups that carry out these simulations publish quantitative
error bars determined by studying the convergence properties of the
solutions as a function of grid size, domain size and other computational
assumptions.  

The remainder of this paper sketches the arguments that lead to these
conclusions.  Detailed derivations and formulae, and applications to
neutron-star simulations, will be the subject of
a future publication.

\section{Orbits at the turning point in post-Newtonian gravity}

In Newtonian gravity, the orbit of a pair of point masses may be
described by the set of equations
\begin{eqnarray}
p/r &=& 1+e \cos(\phi - \omega)  \,, 
\nonumber \\
r^2 \Omega &\equiv& r^2 d \phi /dt = (mp)^{1/2} \,,
\nonumber \\
E &=& \mu ( {\dot r}^2 + r^2 \Omega^2 )/2 - \mu m /r \,,
\nonumber \\
J &=& \mu |{\bf x} \times {\bf v}| \,,
\label{newton}
\end{eqnarray}
where $p = a(1-e^2)$ is the semi-latus rectum ($a$ is the semi-major
axis), $\omega$ is the angle of pericenter, $m=m_1 + m_2$ is the total
mass, $\mu = m_1m_2/m$ is the reduced mass, and $E$ is the total
orbital energy (henceforth we use units in which $G=c=1$).  A circular orbit
corresponds to $e=0$, with $r=a=$ constant, $\Omega^2 = m/a^3$,
$E/\mu = a^2 \Omega^2/2 - m/a = -(m\Omega)^{2/3}/2$, and $J/\mu =
\sqrt{ma} = (m/\Omega)^{1/3}$.  However, if
we  demand only that the orbit be at apocenter, so that $\dot r = 0$
only, we have $\phi = \omega + \pi$, $r = p/(1-e)$, $\Omega^2 = (m/p^3)(1-e)^4$,
and 
\begin{eqnarray}
E/\mu &=&  - {1 \over 2} (1-e^2) \left [ {m\Omega_a \over
(1-e)^2} \right ] ^{2/3} \,, 
\nonumber \\ 
J/\mu m&=& \left [ {m\Omega_a \over (1-e)^2} \right ] ^{-1/3}.
\end{eqnarray}
where $\Omega_a$ is the angular velocity at apocenter.
The energy and angular momentum
expressions when the orbit is at pericenter may be obtained by
letting $\Omega_a \to \Omega_p$ and $e \to -e$. 

At 1PN order, the equations of motion can be solved using either the
direct approach of Wagoner and Will \cite{wagwill}, 
or the ``perturbed osculating
orbit elements approach'' of Lincoln and Will \cite{lincwill} to yield the
orbit equations
\begin{eqnarray}
{\tilde p}/r &=& 1 + {\tilde e} \cos (\phi -\omega) 
\nonumber \\
&& + {\tilde \zeta} \bigl [-(3-\eta)+(1+9\eta/4){\tilde e}^2 
\nonumber \\
&& + (7/2-\eta){\tilde e} \cos (\phi - \omega)
	+ 3{\tilde e} \phi \sin (\phi - \omega) 
\nonumber \\
	&& - (\eta/4){\tilde e}^2 \cos (2\phi -
	2\omega)  \bigr ]  + O({\tilde \zeta})^2 \,,
\nonumber \\
r^2 \Omega &=& (m{\tilde p})^{1/2} [1 - {\tilde \zeta} (4-2\eta) 
{\tilde e} \cos (\phi - \omega)
 	+ O({\tilde \zeta})^2] \,,
\nonumber \\
E &=& \mu ( {\dot r}^2 + r^2 \Omega^2 )/2 - \mu m /r 
\nonumber \\
&& + {3 \over 8} \mu v^4 (1-3\eta) + {1 \over2} \mu (m/r)^2 
+ {1 \over2} \mu \eta {\dot r}^2 m/r  
\nonumber \\
&& 
+ {3 \over 2} \mu (v^2m/r)(1+\eta/3) 
+ O(m/r)^4 \,,
\nonumber \\
J &=& \mu |{\bf x} \times {\bf v}| \bigl \{ 1 + \frac{1}{2} v^2
(1-3\eta) + \frac{m}{r}(3+\eta)  
\nonumber \\
&& \qquad + O(m/r)^2  \bigr \} \,,
\label{1PN}
\end{eqnarray}
where $v^2= {\dot r}^2 + r^2 \Omega^2$, $\eta = \mu/m$, and $\tilde
\zeta = m/\tilde p$.  
The limit ${\tilde e} \to 0$ corresponds to
a circular PN orbit.  However, at higher PN orders, 
neither the orbital eccentricity ${\tilde e}$ nor
the semilatus rectum ${\tilde p}$ 
is uniquely or invariantly defined.  One definition of eccentricity
used by Lincoln and Will \cite{lincwill}  was that of a Newtonian orbit
momentarily tangent to the true orbit (the ``osculating''
eccentricity);  
other authors \cite{ds88} define multiple
``eccentricities'' to encapsulate different aspects of non-circular
orbits at PN order.    

We define an alternative
eccentricity and semilatus
rectum according to:
\begin{eqnarray}
e &\equiv& {{ \sqrt{\Omega_p} - \sqrt{\Omega_a} } \over 
	{ \sqrt{\Omega_p} + \sqrt{\Omega_a} }} \,,
\nonumber \\
\zeta \equiv
{m \over {p}} &\equiv& \left ( {{\sqrt{m\Omega_p} +
\sqrt{m\Omega_a}}
        \over 2} \right )^{4/3} \,,
\end{eqnarray}
where $\Omega_p$ is the value of $\Omega$ where it passes through a
local maximum (pericenter), and $\Omega_a$ is 
the value of $\Omega$ where it passes
through the {\it next}
local minimum (apocenter).  
These definitions have the following virtues: (1) they reduce precisely to
the normal eccentricity $e$ and semilatus rectum $p$
in the Newtonian limit, as can be seen
from Eqs. (\ref{newton}); 
(2) they are constant in the absence of radiation reaction;
(3) they are somewhat more directly connected to measurable quantities, 
since $\Omega$ is the
angular velocity as seen from infinity (eg. as measured in the
gravitational-wave signal) and one calculates only maximum
and minimum values, without concern for the coordinate location in the
orbit; and (4) they are straightforward to calculate in a numerical
simulation of orbits without resorting to complicated definitions of
``distance'' between bodies.  They have the defect that, when
radiation reaction is included, they are not local, continuously evolving
variables, but rather are some kind of orbit-averaged quantities (for
this reason, they may not be as ``covariant'' as they seem -- this
issue is under investigation).  Nevertheless,
when an eccentric orbit decays and circularizes 
under radiation reaction the definition of $e$
has the virtue that it tends naturally to
zero when the orbital frequency turns from ocillatory behavior to
monotonically increasing behavior (i.e. the maxima and minima merge).

By virtue of these definitions, $\zeta$ has the further property that
\begin{equation}
\zeta = \left ( {{m\Omega_p} \over (1+e)^2} \right )^{2/3} 
	= \left ( {{m\Omega_a} \over (1-e)^2} \right )^{2/3} \,.
\end{equation}
It is then simple to show that the relation between $e$ and
$\zeta$ and the
corresponding quantities  used in the Wagoner-Will solution (\ref{1PN})
is 
$e = \tilde e \{ 1 + \tilde \zeta [9/2 - \eta + (1-3\eta){\tilde e}^2] \}$, 
and 
$\zeta = \tilde \zeta \{1- 4 \tilde \zeta [1-\eta/3+(1/3-\eta){\tilde
e}^2  ]\}$.
Applying these definitions to Eqs. (\ref{1PN}), we find 
the 1PN expression for $E(\Omega)$ and $J(\Omega)$
for a non-circular orbit, expressed in terms of $\Omega$ at the
apocenter or pericenter:
\begin{eqnarray}
\frac{E}{\mu} &=& -\frac{1}{2}\left (1- {e}^{2}\right )
\zeta
+\left [
\frac{3}{8}+\frac{\eta}{24}+\left (-\frac{5}{12}+\frac{\eta}{12}\right )
 {e}^{2}  \right .
\nonumber \\
&& \left . +\left (\frac{1}{24}-\frac{\eta}{8}\right ) 
{e}^{4} \right ]
\zeta^2 \,, 
\nonumber \\
\frac{J}{\mu m} &=& \frac{1}{\sqrt{\zeta}} \left \{ 1 + \left [ 
	\frac{3}{2} + \frac{1}{6}\eta- \left ( \frac{1}{6}-
	\frac{1}{2}\eta \right )e^2 \right ] \zeta \right \}
	\,.
\label{EJpn}
\end{eqnarray}
Note that in the circular orbit limit ($e \to 0$), $E$ and $J$ satisfy
the expected relation $dE/d\Omega = \Omega dJ/d\Omega$.  
We have extended this result to 3PN order using the 3PN orbit equations of
Blanchet and Faye \cite{bf01}, 
using both harmonic-gauge and ADM gauge expressions, 
but neglecting radiation-reaction terms at 2.5PN
order (details will be given in a future publication).

For those numerical
simulations in which the bodies are assumed to be corotating, we must
include a number of rotational effects.
First is the energy of rotation of the individual corotating
bodies, 
of order $\delta
E_{\rm rot}/\mu \sim I\Omega^2/\mu \sim (m/r)(R/r)^2$, where $R$ is
the characteristic size of the body.  For compact
bodies, $R \sim M$, so these effects are equivalent to 2PN order and
higher.  
For black holes, we
will use the standard procedure of splitting the mass
of each body into its irreducible mass and its rotational energy using the
Kerr formulae $M=M_{\rm irr}/[1-4(M_{\rm irr}\Omega)^2]^{1/2}$ and
$S=4M_{\rm irr}^3\Omega/[1-4(M_{\rm irr}\Omega)^2]^{1/2}$ \cite{luc02,dgg02}.
Also, since the sequences of numerical simulations hold $m_{\rm irr} =
(M_{\rm irr})_1 + (M_{\rm irr})_2$ fixed
as they vary $\Omega$,
we will expand the masses that appear 
in the Newtonian orbital contribution to the energy and angular momentum about
$m_{\rm irr}$.  These contributions together yield
\begin{eqnarray}
\delta E_{\rm rot}/\mu &=& (2/\eta)(1-3\eta) (m_{\rm irr} \Omega)^2
\nonumber \\
        &&
	+ (6/\eta)(1-5\eta+5\eta^2)(m_{\rm irr} \Omega)^4 
\nonumber \\
	&& -(1-e^2)(\frac{2}{3}-\eta) \zeta (m_{\rm irr} \Omega)^2
\,, 
\nonumber \\
\delta J_{\rm rot}/\mu m_{\rm irr} &=& 
	(4/\eta)(1-3\eta) (m_{\rm irr} \Omega)
\nonumber \\
        &&
        + (8/\eta)(1-5\eta+5\eta^2)(m_{\rm irr} \Omega)^3
\nonumber \\
	&& + \frac{1}{\sqrt{\zeta}} (m_{\rm irr}
\Omega)^2 (\frac{4}{3}-2\eta) \,,
\label{EJrot}
\end{eqnarray}
where henceforth, $\zeta = [m_{\rm irr}\Omega_a/(1-e)^2]^{2/3}$, and
$\mu$ and $\eta$ are expressed in terms of irreducible masses.
The first two terms in each of Eqs. (\ref{EJrot}) are the rotational
terms, expanded to second order, while the third comes from expanding
the Newtonian orbital term.

Similarly spin-orbit
and spin-spin 
effects are of order $\delta E_{\rm S.O.}/\mu \sim LS/\mu r^3$ and
$\delta E_{\rm S.S.}/\mu \sim S_1 S_2 /\mu r^3$
where $L$ denotes the orbital angular momentum and $S$ denotes the
bodies' spin.  For co-rotating bodies, these effects can be shown to
be of order $\delta E_{\rm S.O.}/\mu \sim (m/r)(mR^2/r^3)$
and  $\delta E_{\rm S.S.}/\mu \sim (m/r)(mR^4/r^5)$, which for compact
bodies are equivalent to 3PN and 5PN order, respectively.  Henceforth,
we will ignore the spin-spin terms.  
Assuming co-rotating bodies with spins aligned with
the orbital angular momentum, including both the direct and
orbital effects of the spin-orbit terms \cite{kww,kidderspin}
as well as their effect on our
definitions of $e$ and $\zeta$, we find 
\begin{eqnarray}
{{\delta E_{\rm S.O.}} \over \mu} &=& \left ( - \frac{16}{3}
   	+ 12\eta +  \frac{4}{3} (5-11\eta)e^2
	- \frac{4}{3}(1-2\eta)e^4 \right ) 
\nonumber \\
 && \qquad \times (m_{\rm irr}\Omega) \zeta^{5/2} \,. 
\nonumber \\ 
{{\delta J_{\rm S.O.}} \over {\mu m_{\rm irr}}} 
&=&  \left ( -\frac{40}{3} + 30\eta
	- \frac{2}{3} (4-11\eta)e^2 \right ) (m_{\rm irr} \Omega )
\zeta \,.
\end{eqnarray}

Apart from the generalization to eccentric orbits, these methods parallel
exactly those of Blanchet \cite{luc02}.

For tidal interactions, the contribution to the orbital
energy is of order $\delta E_{\rm tidal}/\mu \sim (m/r)(R/r)^5$.  
For black holes, tidal effects are thus equivalent to 5PN
order, and will be neglected.  For neutron stars, $R \sim 5M$, and
tidal effects must be included, however it is sufficient to use
Newtonian gravity to calculate them.  These will be discussed elsewhere.  

\begin{figure}
\leavevmode
\psfig{figure=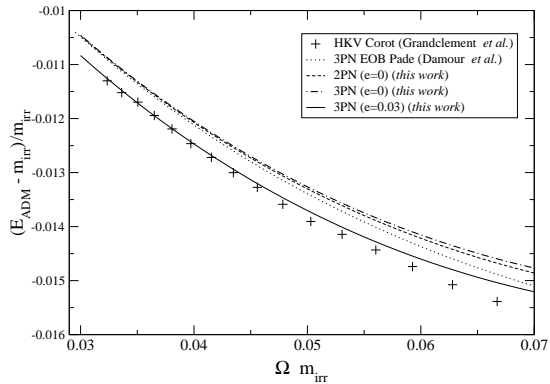,angle=270,width=8cm}
\caption{\label{Fig1}Binding energy of equal-mass, co-rotating black holes
vs. angular frequency.}
\end{figure}

\begin{figure}
\leavevmode
\psfig{figure=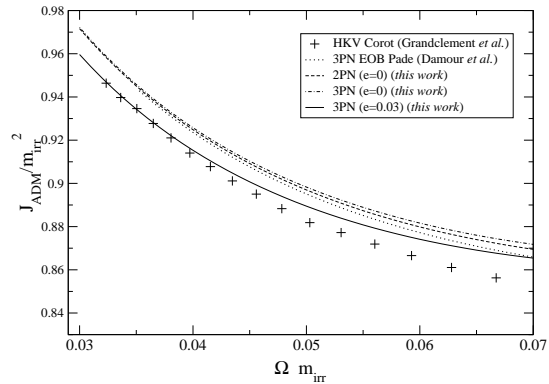,angle=270,width=8cm}
\caption{\label{Fig2}Angular momentum of equal-mass, co-rotating black
holes
vs. angular frequency.}
\end{figure}

\section{Comparison with numerical simulations of
corotating binary black holes}

We combine the PN, rotational and spin-orbit contributions to $E$ and $J$,
set $\Omega_a = \Omega$ so that the stars corotate at the orbital
angular frequency at apocenter, set $\eta = 1/4$ for equal-mass black
holes, 
and plot
$(E - m_{\rm irr})/m_{\rm irr}$ and $J/m_{\rm irr}^2$ versus $m_{\rm irr}
\Omega$,  
where 
$m_{\rm irr}$ denotes the total irreducible mass.  The results are shown in
Figures 1 and 2.  For $e=0$, we show results both from 
the 3PN orbital expressions,
and from truncated 2PN orbital expressions, while for $e=0.03$ we plot the
full 3PN results.  The
crosses denote the corresponding 
numerical data of Grandcl\'ement et al.\cite{ggb02}; the other curve is the
3PN resummation result of Damour et al.\cite{dgg02} for circular orbits.  We
note that all circular-orbit PN estimates, including
conventional 2PN and 3PN approaches \cite{luc02}, Pad\'e and resummation
approaches \cite{dgg02}, and our approach, agree well among
themselves, but are systematically displaced from the numerical
points, both for $E$ and for $J$.   This can also be seen in Figures 5
and 6 of \cite{dgg02}. 
By contrast,  with $e \approx 0.03$, our PN estimate fits the curves
perfectly for small values of  $m_{\rm irr} \Omega$. 
This fit is merely illustrative: if the numerical simulations
do have some eccentricity, there is no reason why the eccentricity
should be the same for each numerical point.  Figure 3 shows the values of
$e$ that best fit the numerical data over the range of $\Omega$ considered.  

\begin{figure}
\leavevmode
\psfig{figure=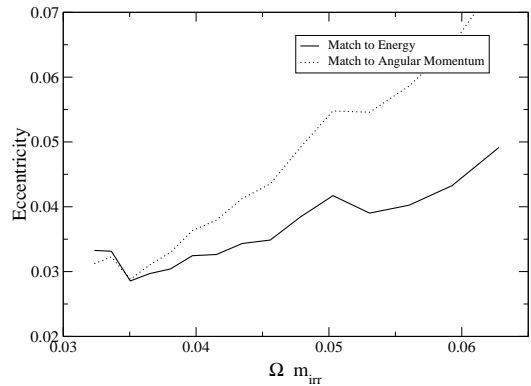,angle=270,width=8cm}
\caption{\label{Fig3}Values of eccentricity giving a match to numerical
HKV simulations.}
\end{figure}
The conclusions are not substantially different if we use
the Wagoner-Will type eccentricity $\tilde e$. 
Whatever definition one uses, $e=0$ gives
a worse fit than a finite eccentricity.
 
\section{Discussion}

High-order post-Newtonian approximations for circular orbits
appear to give results for
the energy and angular momentum of corotating binary black holes well
away from the innermost orbit that
are in excellent agreement -- with each other.  This is illustrated in
Figure 4, which plots the energy from four PN results quoted by Damour et al
\cite{dgg02}, and from our 2PN and 3PN results, in the range around $\Omega
m_{\rm irr} \sim 0.32$.  Apart from the one 2PN
result quoted by Damour et al., all are within 0.5 percent of each
other.  This could be viewed as an estimate of the accuracy of the PN
approximation in this regime.  But all results are four percent displaced
from the numerical HKV data.  In the absence of a quantitative estimate of the
accuracy of the numerical simulations in this regime, it is difficult
to decide if this difference is a signal of a physical effect, such as
the small eccentricity suggested by our work.  

\begin{figure}
\leavevmode
\psfig{figure=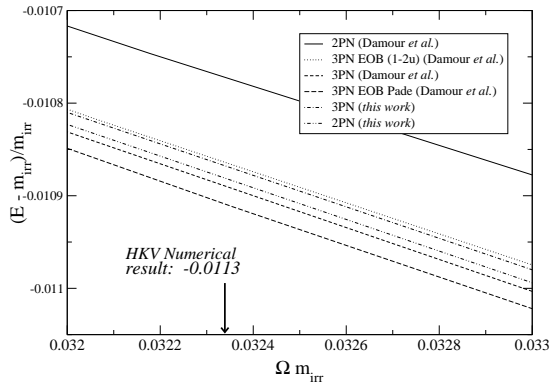,angle=270,width=8cm}
\caption{\label{Fig4}Comparison of PN calculations of energy for widely
separated black holes.}
\end{figure}

One way to test whether the orbits represented by the HKV numerical
simulations are really eccentric would be to evolve the orbits
numerically for a short period of time, say 1/4 of an orbit ($\delta \phi =
\pi/2$).  For a nearly circular orbit of equal mass bodies,
gravitational radiation reaction
should cause the parameter $\zeta = m/p = (m\Omega)^{2/3}$ to evolve
according to $d\zeta/ d\phi = (16/5) \zeta^{7/2}$ (see, for example
Eq. (3.11a) of \cite{lincwill}).  Thus, over a
quarter of an orbit, $\zeta$ should change by approximately $\delta
\zeta /\zeta \approx (8\pi/5) \zeta^{5/2} \approx (8\pi/5)
(m\Omega)^{5/3} \approx 0.016$, for $m\Omega \sim 0.032$.  Hence the
orbital separation should decrease by about 1.6 percent or the angular
velocity should increase by 2.4 percent.  By contrast, if the orbit has
an eccentricity of 0.03 and is at apocenter,
then in a quarter of an orbit, the separation
should decrease by 3 percent, while the angular velocity should
increase by 6 percent.  Even with radiation damping, the orbit should
pass through a distinct pericenter when $\delta \phi \approx \pi$, 
and the orbital separation should
increase, while the angular velocity decreases.

\begin{acknowledgments}

We are grateful to Luc Blanchet,  Mark Miller and Wai-Mo Suen for useful
discussions, and to Eric Gourgoulhon and Thibault Damour for providing the
numerical and PN data from Figs. 5 and 6 of \cite{dgg02} for our comparisons.
This work was supported in part by the National Science Foundation under
grant No. PHY 00-96522.  T.M. was supported in part by an internship from 
the Centre Nationale de la Recherche
Scientifique.
\end{acknowledgments}

\end{document}